\documentclass[usenatbib]{mn2e}
\input{epsf}
\usepackage{amssymb}
\usepackage{natbib}
\usepackage{color}
\usepackage{journals}

\newbox\grsign \setbox\grsign=\hbox{$>$} \newdimen\grdimen \grdimen=\ht\grsign
\newbox\simlessbox \newbox\simgreatbox
\setbox\simgreatbox=\hbox{\raise.5ex\hbox{$>$}\llap
     {\lower.5ex\hbox{$\sim$}}}\ht1=\grdimen\dp1=0pt
\setbox\simlessbox=\hbox{\raise.5ex\hbox{$<$}\llap 
     {\lower.5ex\hbox{$\sim$}}}\ht2=\grdimen\dp2=0pt

\newcommand{\hMpc}{{\ifmmode{h^{-1}{\rm Mpc}}\else{$h^{-1}$Mpc }\fi}}
\newcommand{\hkpc}{{\ifmmode{h^{-1}{\rm kpc}}\else{$h^{-1}$kpc }\fi}}
\newcommand{\hMsun}{{\ifmmode{h^{-1}{\rm {M_{\odot}}}}\else{$h^{-1}{\rm{M_{\odot}}}$}\fi}}
\newcommand{\Msun}{{\ifmmode{{\rm {M_{\odot}}}}\else{${\rm{M_{\odot}}}$}\fi}}

\title[Orbital anisotropy]{Orbital anisotropy in cosmological haloes
  revisited} \author[R. Wojtak]{Rados{\l}aw Wojtak,$^{1}$ Stefan
  Gottl\"ober $^{2}$ \& Anatoly Klypin$^{3}$
  \\   \\
  $^1$Dark Cosmology Centre, Niels Bohr Institute, University of
  Copenhagen, Juliane Maries Vej 30, DK-2100 Copenhagen \O,
  Denmark\\
  $^2$Leibniz-Institute f\"ur Astrophysik Potsdam (AIP), An der Sternwarte 16, 14482 Potsdam, Germany\\
  $^3$Astronomy Department, New Mexico State University, Las Cruces, NM 88003, USA\\
}

\begin{document}

\maketitle

\begin{abstract}
  The velocity anisotropy of particles inside dark matter (DM) haloes is an
  important physical quantity, which is required for the accurate modelling of mass
  profiles of galaxies and clusters of galaxies. It is typically
  measured using the ratio of the radial-to-tangential velocity
  dispersions at a given distance from the halo centre.  However, this
  measure is insufficient to describe the dynamics of realistic haloes,
  which are not spherical and are typically quite elongated. Studying
  the velocity distribution in massive DM haloes in cosmological
  simulations, we find that in the inner parts of the haloes the local
  velocity ellipsoids are strongly aligned with the major axis of the
  halo, the alignment being stronger for more relaxed
  haloes.  In the outer regions of the haloes, the alignment becomes
  gradually weaker and the orientation is more random. 
 These two distinct regions of different degree of the alignment coincide with
 two characteristic regimes of the DM density profile: a shallow
 inner cusp and a steep outer profile that are separated by
  a characteristic radius at which the density declines as $\rho\propto
  r^{-2}$. This alignment of the local velocity ellipsoids requires
  reinterpretation of features found in 
  measurements based on the spherically averaged ratio of the
  radial-to-tangential velocity dispersions. In particular, we show
  that the velocity distribution in the central halo regions is highly
  anisotropic. For cluster-size haloes with mass
  $10^{14}-10^{15}\hMsun$, the velocity anisotropy along the major axis
  is nearly independent of radius and is equal to $\beta
  =1-\sigma^2_{\rm perp}/\sigma^2_{\rm radial}\approx 0.4$, which is
  significantly larger than the previously estimated spherically averaged
  velocity anisotropy. The alignment of density and velocity anisotropies, and the
  radial trends may also have some implications for the mass modelling
  based on kinematical data of such objects as galaxy clusters or
  dwarf spheroidals, where the orbital anisotropy is a key element in
  an unbiased mass inference.

\end{abstract}


\section{Introduction}

The orbital anisotropy describes the distribution of orbits in
astrophysical systems. It plays a key role in dynamical modelling of
kinematical data of objects at all scales, from the stellar halo of 
the Milky Way \citep{Kaf12}, dwarf spheroidals
\citep{Lok09,Wal09}, to elliptical galaxies
\citep{Dek05,Nap11,Woj12} and galaxy clusters
\citep{Biv03,Woj10}. Prior knowledge on the anisotropy or elaborated
techniques of data analysis are essential for accurate and unbiased
mass estimates. The main difficulty arises from the well-known
mass-anisotropy degeneracy occurring in the Jeans analysis of the
velocity dispersion profiles \citep{Mer87}. Several methods were
developed to break this degeneracy 
\citep[see e.g.][]{Lok02,Lok03,Woj09,Wol10,Mam12}. Their efficiency,
however, critically relies on the quality of the data. In addition,
 observational constraints on the mass
profiles are still affected by what dynamical models assume about the
anisotropy or what priors on the anisotropy are used. As the matter of
fact, most models rely on certain parametrisation of the anisotropy
\citep[e.g.][]{Lok02,Mam12,Woj09} or incorporate some well-motivated
profiles \citep[e.g.][]{Dia99,Dek05} or prior distributions for its
parameters \citep[e.g.][]{New12}. The choice of the prior probability
and parametrisation is commonly motivated by cosmological $N$-body
simulations. The orbital anisotropy in DM haloes is often used as a
point of reference in such preselection of dynamical models. The best
example showing this effect of feedback from the simulations is the 
commonly adopted assumption that the velocity distribution in the
centre of gravitationally bound cosmological objects is isotropic or
nearly isotropic \citep[see e.g.][as examples in studies of dwarf
spheroidals and galaxy clusters]{Wal06,New12}.

The orbital anisotropy is also a quantity of great interest 
in the context of studying velocity distributions of DM particles in simulated haloes. 
Many studies showed a number
of interesting properties such as a relation between the anisotropy
and DM density profile \citep{Han06,Zai08} or existence of a universal
attractor in the space spanned by all solutions of the Jeans equation
\citep{Han10}. The anisotropy parameter is 
often a part of equilibrium models describing DM haloes \citep[see e.g.][]{Deh05}, 
though recent studies showed that it is not relevant in mapping between 
the mass and the phase-space density profiles \citep{Lud11}. 
A number of studies addressed such problems as the
radial profiles of the anisotropy \citep{Woj05,Asc08}, the bias
between DM particles and subhaloes \citep{Die04}, evolution of the
anisotropy in controlled simulations of halo mergers \citep{Spa12},
the redshift evolution \citep{Ian12} and dependance on the mass and
dynamical equilibrium of DM haloes \citep{Lem12}. The overall picture
emerging from these studies points to the fact that the velocity
distribution is nearly isotropic in the halo centre and radially
biased at large radii (excess of radial orbits). Although this trend
is common to all DM haloes, the anisotropy profiles of individual
haloes are significantly scattered around the mean trend. Deviation
from the mean trend seems to depend on dynamical stage of the haloes
\citep{Lem12}.

The global velocity ellipsoids of DM haloes are aligned with the major
axes of the halo shape \citep{Kas05,Allgood,Sar12}. This property seems to
occur also locally, although this problem has not been extensively
studied in the literature \citep{Zem09}. Interestingly, similar
configuration of the velocity ellipsoid is consistent with the
kinematical data of elliptical galaxies \citep{Cap07} and galaxy
clusters \citep{Ski12} what suggests that it is probably a generic
feature of the orbital structure in cosmological objects. Despite
these facts, the orbital anisotropy in simulated DM haloes is commonly
quantified in terms of the ratio of the radial-to-tangential velocity
dispersions which breaks the preferred symmetry of the velocity
distributions. This raises the question to what extent the orbital
anisotropy defined in this way describes the true distribution of
orbits in DM haloes.

The anisotropy of the velocity dispersion tensor is commonly
quantified by the anisotropy parameter \citep{Bin08}
\begin{equation}
\beta=1-\frac{\sigma_{t}^{2}}{2\sigma_{r}^{2}},
\label{beta}
\end{equation}
where $\sigma^2_{\rm r}$ and $\sigma^2_{\rm t}$ are the  radial and
tangential velocity dispersions, respectively. This is a reasonable
definition for a spherical or almost spherical halo because in this
case the velocity dispersion tensor should be oriented along the
radial direction and, because of the symmetry, one expects that
dispersions perpendicular to the radius are equal:
$\sigma^2_{\theta}=\sigma^2_{\phi}= \sigma^2_{t}/2$. For haloes with
substantial elongation -- and a large fraction of cluster-size haloes
are elongated -- the velocity ellipsoid is more complex and simple
spherically averaged $\beta(r)$ is not enough to describe such systems.

Despite obvious inconsistencies between the spherically averaged
$\beta$ and the elongation of DM haloes, the anisotropy
parameter $\beta$ is commonly used to study the phase-space properties
of DM haloes. In realistic cases one expects that local velocity
tensor is neither aligned along radius nor tangential velocity dispersions 
are equal. To complicate the situation, the velocity dispersions are also
expected to depend on all three space coordinates, not just one
radius. So, realistic $\beta(\vec r)$ is a complex entity. To some
degree, this justifies using spherically averaged $\beta$. Realistic
$\beta(\vec r)$ is not well studied and, even it were, it would be more
difficult to adopt it in dynamical models. Instead, by using spherically averaged
$\beta$, one gets a practical tool to build dynamical models with
understanding that there is a price for this simplification: one should
expect some errors in the models. Here we do not study the errors and
focus only on description of the realistic velocity ellipsoids.

In this paper, we study the properties of the velocity ellipsoids in
cluster-size DM haloes in high-resolution cosmological simulations.
The paper is organised as follows. In section 2, we describe
cosmological simulations and the properties of DM haloes used for the
analysis. In section 3, we investigate the local values of the anisotropy
parameter defined using the radial and tangential velocity dispersions. This
gives us insights into dynamical structure of the halo and it shows
that anisotropy parameter depends on the position with respect to
the halo major axis. This ``classical'' treatment of the anisotropy
parameter neglects the fact that velocity ellipsoid may not be aligned
along radius and that it may be triaxial.  In section 4, we study the
properties of the local velocity dispersion tensor, e.g. alignment,
triaxiality and anisotropy, and we show that the local velocity
ellipsoids are aligned with the major axis of the halo shape.  In
section 5, we calculate spherically averaged (measured in spherical
shells) profiles of the orbital anisotropy defined in the symmetry
consistent with the alignment of the local velocity ellipsoids and
compare with the local values. We also show how the anisotropy
parameter based on the ratio of the radial-to-tangential velocity
dispersions misrepresents the true orbital structure in DM
haloes. Section 6 contains discussion and conclusions. In this 
section, we also provide a simple analytical model for the local velocity 
ellipsoids measured in DM haloes.

\section{Method}
\subsection{Simulations}

\begin{figure*}
\begin{center}
    \leavevmode
    \epsfxsize=16cm
    \epsfbox[50 50 600 640]{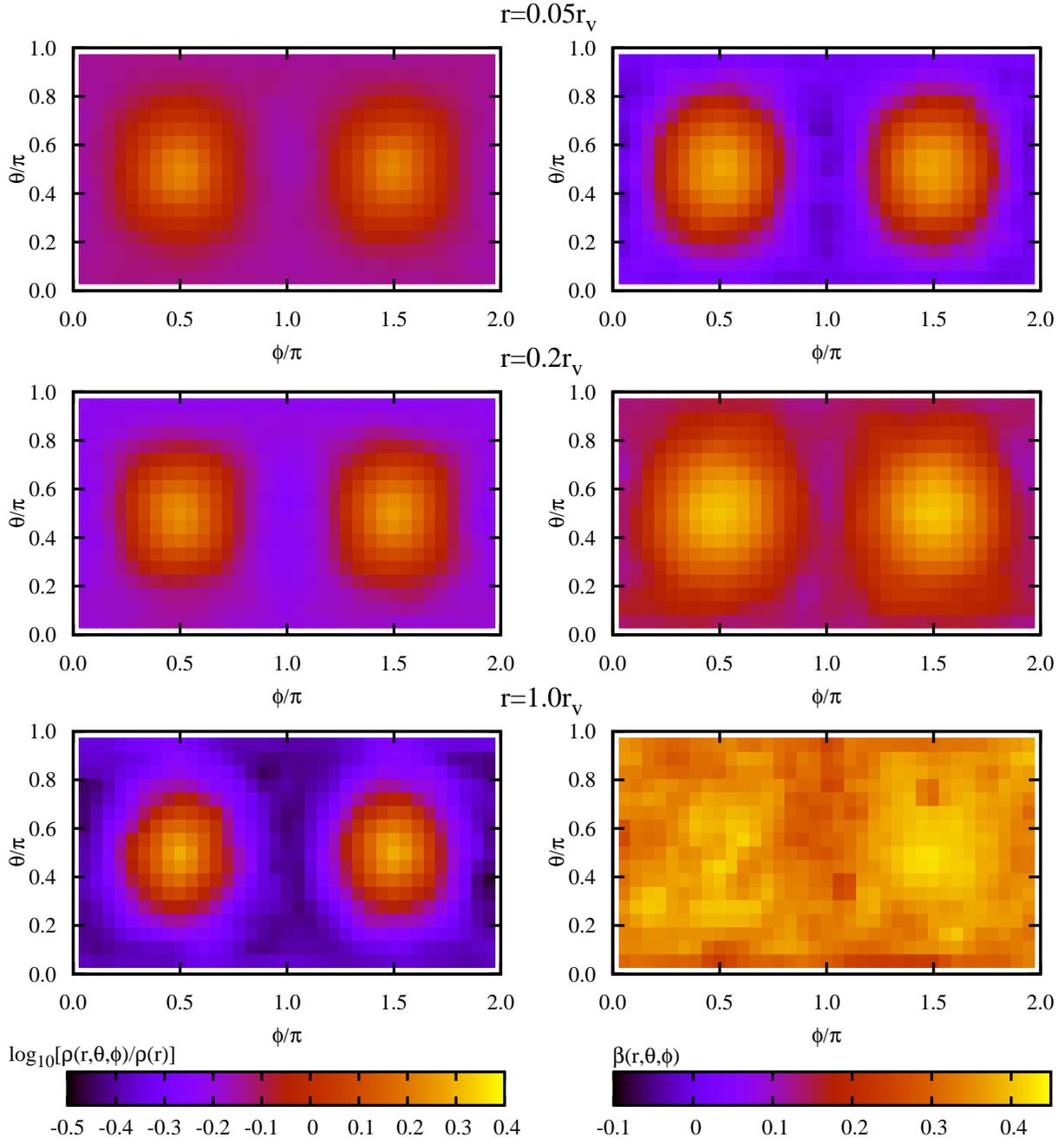}
\end{center}
\caption{Density and velocity anisotropy of DM haloes as a function of
  the polar and azimuthal angles $\theta$ and $\phi$. The system of
  coordinates is chosen in such a way that the major axis is 
  placed at $\theta=\pi/2$ and $\phi=\pi\pm\pi/2$. 
  {\it Left panels:} Deviations of the local density from
  the average density at given radius $\rho(r,\theta,\phi)/\rho(r)$.
  {\it Right panels:} Local velocity anisotropy parameter $\beta$. The
  rows show the maps calculated inside three spherical shells of radii
  equal to $0.05$, $0.20$ and $1.00$ of the virial radius.  When
  measured in spherical shells the density anisotropy declines with 
  decreasing radius. However, the pattern of elongation along the
  major axis is clearly preserved at all radii. The velocity
  anisotropy $\beta(r,\theta,\phi)$ behaves very differently as
  compared with the density $\rho(r,\theta,\phi)$. Close to the virial
  radius the velocities are predominantly radial in all directions:
  $\beta\approx 0.4$. At smaller distances $\beta$ has same value
  $\sim 0.4$ along the major axis and gets to $\beta\lesssim 0$ in the plain
  perpendicular to the major axis.}
\label{beta-density}
\end{figure*}

We use cluster-size DM haloes
selected from the Bolshoi simulation \footnote{The simulation is
  publicly available through the MultiDark database
  (http://www.multidark.org). See \citet{Rie11} for all details of the
  database.} \citep{Kly11}. The simulation follows the evolution of
DM structures in the framework defined by a $\Lambda$CDM
cosmological model with cosmological parameters consistent with recent
measurements based on WMAP five-year data release \citep{Kom09} and
abundance of the Sloan Digital Sky Survey clusters \citep{Roz10}. The
simulation box has a side length equal to $250\hMpc$ and contains $2048^{3}$
particles, each with mass of $1.35\times10^{8}\hMsun$. High mass
resolution of the simulation is a key feature to study the properties
of the local velocity distributions in DM haloes. For all details on the 
simulations, we refer the reader to \citet{Kly11}.

DM haloes are found using the Bound-Density-Maxima (BDM) algorithm
\citep{Kly97}. BDM finds all density maxima with density estimated with the top-hat
filter containing 20 particles. Among all density maxima inside a
given distinct halo the code finds the one which has the deepest
gravitational potential and uses it as the centre of the halo. The
halo virial mass $M_{\rm v}$ is defined as a spherical overdensity
mass with the mean density $\Delta=3M_{\rm v}/(4\pi r_{\rm
  v}^{3})=97.2$ times greater than the critical density, where $r_{\rm
  v}$ is the virial radius. For our study, we select all $517$ haloes
with virial masses greater than $7\times10^{13}\hMsun$. The selected haloes
contain from $5\times10^{5}$ to $8\times10^{6}$ particles inside the
virial sphere. Velocities of DM particles are corrected for the halo bulk 
velocities approximated by the mean velocity of the particles inside 
the virial sphere.

\subsection{Halo shapes}

We quantify the shape of the haloes in terms of the shape tensor given by
\begin{equation}
I_{i,j}=\sum_{n=1}^{N}r_{i,n}r_{j,n},
\label{shape-tensor}
\end{equation}
where $r_{i,n}$ is $i$-th component of the position vector with
respect to the halo centre and the sum is over all particles lying
inside the virial sphere. Many authors use the normalised positions 
in order to minimise effect of substructures or to equalise contribution from 
the particles at small and large distances. However, as shown by \citet{Zem11}, 
this weighting leads to a bias when one measures the global shape of the halo. 
In this case, assigning equal weights to all particles is recommended.

Eigenvectors of the tensor (\ref{shape-tensor}) determine the semi-principle axes of an ellipsoid 
approximating the halo global shape. Information which is essential for our study is the eigenvector
associated with the major axis of the halo shape ellipsoid  -- the axis
minimising the moment of inertia.

In general, the semi-principle axes of the halo shape ellipsoid may depend on radius. 
Cosmological simulated haloes, however, exhibit strong coherence of orientations of the shape ellipsoids 
measured at different radii \citep[][]{Jin02,Bai05}. This means that the major axis of the global halo shape is 
a well-defined global axis of symmetry, both in the inner and outer parts of the haloes. 
We use this axis as a reference axis for studying symmetry of the velocity distributions 
in the selected haloes.

\subsection{Relaxed haloes}

In our analysis, we also consider a subsample of relaxed haloes. Our
criteria of relaxedness are similar to those proposed by \cite{Net07}
and are based on three diagnostics of dynamical equilibrium: the
offset between the mass centre and the minimum of the potential, the
offset between the bulk velocity of the halo and the mean velocity of
the most gravitationally bound particles, and the virial ratio. We use
BDM halo centres as the positions of the minimum of the potential and
BDM bulk velocities as velocities of the most gravitationally bound
particles. The virial ratio is estimated using $3\times10^{3}$
particles randomly drawn from every halo. The gravitational binding
energy is computed using direct summation.

We define relaxed haloes as those which satisfy three conditions: the
offset of the mass centre less than $0.07r_{v}$
and the virial ratio less than
$1.35$. The limits imposed on all diagnostics separate the outliers
which populate long tails of the distributions and are associated with
unrelaxed haloes. Most unrelaxed haloes ($76$ per cent) are found by
the offset of the mass centre. Our selection criteria yield $299$
relaxed haloes ($58$ per cent of the total number).

\section{Velocity anisotropy along and perpendicular to the major axis}
We start our analysis by investigating the global trends in the
velocity anisotropy. Here we ask a simple question: how does $\beta$ change 
along and perpendicular to the major axis of the density distribution.

We compute the local values of the anisotropy parameter on a regular
grid of spherical angles inside spherical shells. The polar $\theta$
and azimuthal $\phi$ angles of the halo major axis are set
consistently at the same values in all haloes (for the sake of
readability of the plots, we chose $\theta=\pi/2$ and
$\phi=\pi\pm\pi/2$).  We use three spherical shells of
radii $(0.03-0.07)r_{\rm v}$, $(0.16-0.25)r_{\rm v}$ and
$(0.93-1.07)r_{\rm v}$ containing on average $3.9\times10^{4}$,
$1.2\times10^{5}$ and $10^{5}$ particles. The
anisotropy parameter is calculated inside the cones with opening angle
of $10$ degrees around every point of the regular grid representing 
a cylindrical map projection of a sphere (the Mercator projection), for every
spherical shell. At every grid point, we find the median value of the
anisotropy parameter in the halo sample. In all results presented
below we show the medial values.

\begin{figure}
\begin{center}
    \leavevmode
    \epsfxsize=8cm
    \epsfbox[50 50 600 420]{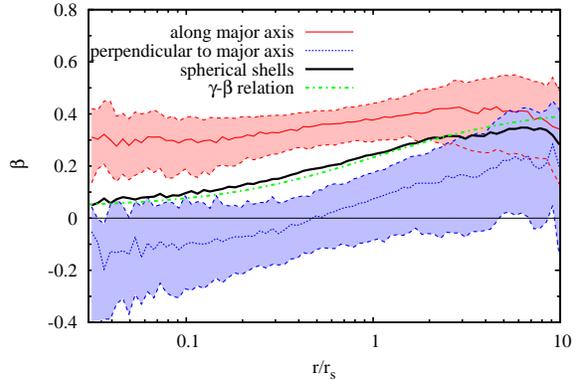}
\end{center}
\caption{Radial profiles of the anisotropy parameter $\beta$. 
The profiles are complementary to Figure~\ref{beta-density} 
showing dependance of $\beta$ on the position with respect to the halo major axis. 
Spherically averaged $\beta(r)$ indicates preferentially
  radial anisotropy at all radii. It also slightly increases with radius, as
  found in many studies. Instead, the anisotropy along the density major axis
  is nearly constant at all radii (with $\beta\approx 0.35$), whereas regions
  perpendicular to the major axis show tendency to have 
  excess of the tangential velocity dispersion 
  at small distances ($\beta\lesssim 0$). The black dashed line is 
  the linear relation between the anisotropy and the logarithmic density slope from \citet{Han06}. 
  Shaded regions on show $50$ per cent statistical uncertainties.}
\label{beta-maj-min}
\end{figure}

  The right panels of
Figure~\ref{beta-density} show the dependance of the anisotropy
parameter $\beta$ on the position with respect to the
halo major axis. The left panels show the values of the ratio
$\rho(r,\theta,\phi)/\rho(r)$, where $\rho(r,\theta,\phi)$ is the
density inside the intersection of a radial shell and a cone directed
to $(\theta,\phi)$, and $\rho(r)$ is the density inside a shell.

The maps of the local anisotropy parameter inside the two inner shells
show a dipole-like dependance on the position with respect to the halo
major axis. The value of $\beta$ changes from $\approx0.3$ 
along the major axis to $\approx 0$ in the plane 
perpendicular to the major axis. Similar
dipole structure, although not as prominent as in the inner shells, is
also visible at the virial sphere.  Here, the local anisotropy
parameter tends to change from $0.5$ (along major axis) to $0.3$ (in the
perpendicular plane). We note that the same spatial variation of the local
anisotropy parameter has been recently found in non-cosmological
controlled simulations of mergers of DM haloes \citep{Spa12}. In every
shell, the dipole of the anisotropy parameter coincides with the
dipole of the density ratio (compare the left and right panels in
Figure~\ref{beta-density}).

The values of the spherically averaged velocity anisotropy (computed
in spherical shells) are $0.14$, $0.20$ and $0.30$, from the innermost
to outermost shell respectively. One can immediately see 
that the local values of $\beta$ are substantially larger along the major 
axis and smaller in the perpendicular plane compared to the spherically 
averaged values. The local values of $\beta$ in the two innermost shells 
are consistent with similar measurements obtained by \citet{Zem09} 
from high resolution simulations of a Milky Way size DM halo.

Figure~\ref{beta-maj-min} shows the radial profiles of the orbital anisotropy 
along (up to $30$ degrees from the axis) and perpendicular (up to $\pm 20$ degrees from the plane) 
to the major axes of the haloes. We scaled radii by the characteristic 
radius $r_{\rm s}$ where the density profile is proportional to $r^{-2}$. The scale 
radii $r_{s}$ were obtained from fitting the NFW profile \citep{Nav97} to the density 
measured in spherical shells of radii equally spaced in logarithmic scale. 
The quartiles of the anisotropy 
parameter at each radius were calculated using only these radial bins which lie 
inside the virial sphere (the mean virial radius of the haloes is equal to $6.2r_{-2}$). 
The anisotropy along the halo major axes is 
substantially larger than that the spherically averaged one. Its radial dependance is much 
weaker then the profile corresponding to the linear relation between the anisotropy and 
the logarithmic slope of the density profile from \citet{Han06}. The anisotropy in the plane 
perpendicular to the major axes are substantially smaller and negative and small radii. 
The profile is consistent with a linear density slope-anisotropy relation 
with the slope similar to that from \citep{Han06}, but different intercept.

 Alignment of the velocity
anisotropy dipole with the density dipole suggests that the apparent
dependance of $\beta$ on the position with respect to the halo major
axis may be a purely geometrical effect resulting from a wrong
assumption that the local velocity ellipsoids are aligned with the
radial directions (assumption underlying the definition of
$\beta$). In order to address this problem in more detail, we shall
study the properties of the local velocity dispersion tensors.

\section{Local velocity ellipsoids}
The anisotropy parameter $\beta=1-\sigma_{t}^{2}/(2\sigma_{r}^2)$ measures 
the true anisotropy of the velocity distribution only in spherical systems, where one 
can assume that velocity dispersion tensor is diagonal in spherical coordinate 
system and two diagonal coefficients are equal: 
$\sigma_{\theta}^{2}=\sigma_{\phi}^{2}=\sigma_{t}^{2}/2$. In non-spherical systems with distinct 
centres such as DM haloes, $\beta$ is just a function of $6$ non-vanishing coefficients of the 
velocity dispersion tensor. This makes this parameter incapable of disentangling 
the anisotropy of the velocity distribution from its spatial orientation (shape from 
orientation of the local velocity ellipsoids -- geometrical representations of the velocity 
dispersion tensor). In particular, $\beta=0$ may result from non-radially oriented 
and elongated velocity ellipsoids.

In this section, we measure all components of the velocity dispersion 
tensor at different positions in selected DM haloes. Complete information on the tensor 
allows to disentangle the shape from the orientation of the local velocity ellipsoids. 
The local velocity dispersion tensor is calculated in a volume element around 
a certain position as
\begin{equation}
T_{i,j}=\frac{1}{N-1}\sum_{n=1}^{N}v_{i,n}v_{j,n}
\label{sigma-tensor}
\end{equation}
where $v_{i}$ are the velocity components of the particles with
respect to the mean velocity of the volume element and $N$ is the
number of DM particles inside the volume element.

We compute the velocity dispersion tensor in several spherical shells
on a grid determined by sphere pixelisation defined by the HEALPix
code \citep{Gor05}, with the total number of pixels fixed at $48$. The
polar angle of the pixels is measured with respect to the halo major
axis. This allows for an easy selection of the pixels at equal angular
distances from the halo major axis. We consider two subsamples of the
pixels: $8$ pixels around the major axis with $|\cos\theta|>2/3$ and $8$
pixels in the plane perpendicular to the major axis, with $|\cos\theta|<1/3$.

We select DM particles using six spherical shells of radii: 
$(0.00-0.04)r_{\rm v}$, $(0.04-0.08)r_{\rm v}$, $(0.08-0.12)r_{\rm v}$, 
$(0.19-0.21)r_{\rm v}$, $(0.48-0.52)r_{\rm v}$ and
$(0.98-1.02)r_{\rm v}$.  This choice of the shells leads to approximately 
$10^{3}$ particles per pixel. We checked that results presented in 
this section remain the same when recalculated with smaller random subsamples 
of the particles. This ensures us that the number of the particles is sufficient to 
obtain unbiased statistical properties of the 
velocity dispersion tensor.

We calculate all six independent components of the velocity dispersion
tensor using the eq.~(\ref{sigma-tensor}). The calculation is repeated in 
every pixel of all radial bins and the haloes. Then, we diagonalise every tensor 
obtaining three eigenvalues $\sigma^{2}_{1}>\sigma^{2}_{2}>\sigma^{2}_{3}$ 
and three associated eigenvectors. We calculate the distributions of various properties of the local
velocity dispersion tensor in the halo sample. The distributions are
computed by combining information from the same sets of pixels in all
DM haloes. Therefore, the resulting distributions represent
pixel-weighted statistics (all haloes contribute equally to the
distributions). We consider several combinations of the pixel and halo
subsamples: all $48$ pixels from all haloes ($48\times517$ data
points), $8$ pixels along the major axis (or in the perpendicular plane) from all haloes ($8\times 517$
data points), $48$ pixels from relaxed haloes ($48\times267$ data
points).

\subsection{Prolateness}

Geometrical representation of the velocity dispersion tensor is a
triaxial ellipsoid with the axes proportional to $\sigma_{1}$,
$\sigma_{2}$ and $\sigma_{3}$. The shape of such ellipsoid
may be characterised by the triaxiality parameter
\begin{equation}
Z=\frac{\sigma_{1}-\sigma_{2}}{\sigma_{2}-\sigma_{3}}
\label{tri}
\end{equation}
introduced by \citet{Bin85} for description of the shape in the
position space and adopted here for the velocity space. This parameter
measures the degree of prolateness/oblateness of the ellipsoids. Two
limiting cases with $Z=0$ and $Z=1$ correspond to oblate and prolate
ellipsoids, respectively.

\begin{figure}
\begin{center}
    \leavevmode
    \epsfxsize=8cm
    \epsfbox[67 60 645 668]{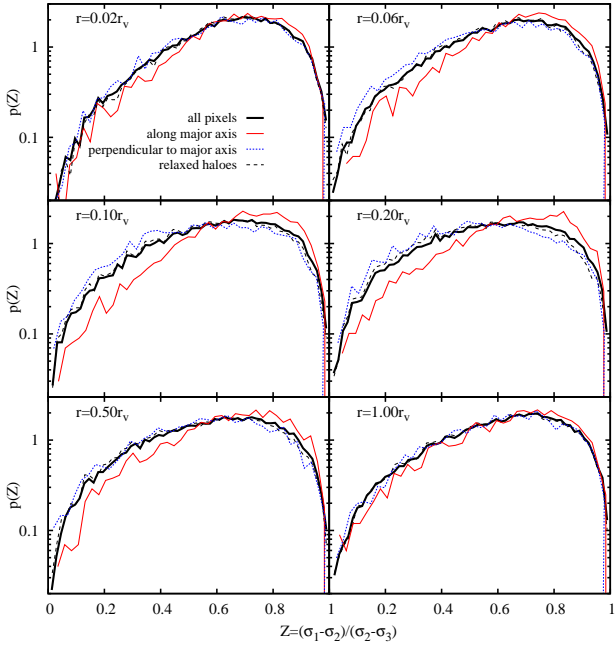}
\end{center}
\caption{Triaxiality of the local velocity ellipsoids in DM
  haloes. The profiles show the distributions of the triaxiality
  parameter defined by the eq.~(\ref{tri}). Velocity ellipsoids are predominantly 
  prolate at all radii. The six panels show the
  distributions in $6$ spherical shells of radii as indicated in the
  left upper corners. The black (thick solid), red (solid) and blue
  (dotted) lines show the profiles for different pixel selection: all
  $48$ pixels, $8$ pixels along the halo major axis and $8$ pixels 
  in the plane perpendicular to the major axis, respectively. 
  The black dashed profiles are the distributions inside
  relaxed haloes.}
\label{sig-pro-pdf}
\end{figure}

Figure~\ref{sig-pro-pdf} shows the distributions of the triaxiality
parameter $Z$ of the local velocity ellipsoids. The ellipsoids tend to
be preferentially prolate at all radii of the haloes.  The most
probable $Z$ equals to $0.7$ and does not vary with radius. The
velocity ellipsoids along the halo major axis appear to be slightly
more prolate than those in the perpendicular plane. This difference
disappears in the innermost shell.

\subsection{Alignment}

\begin{figure}
\begin{center}
    \leavevmode
    \epsfxsize=8cm
    \epsfbox[67 60 645 668]{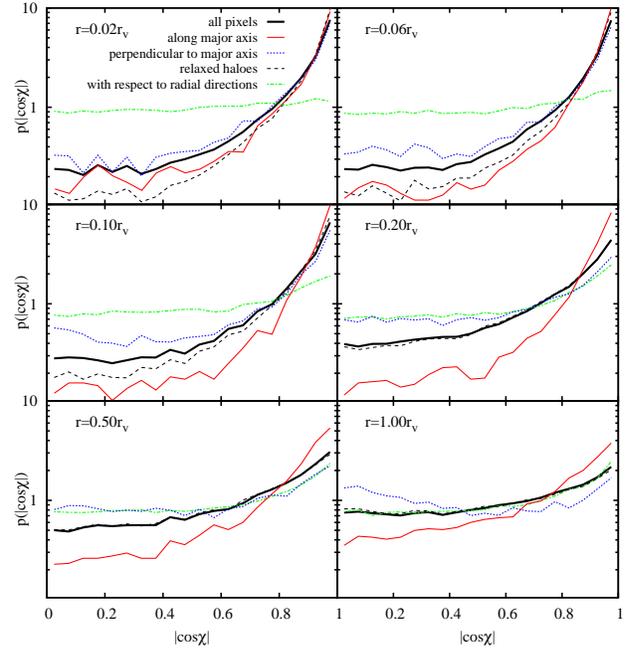}
\end{center}
\caption{Alignment of the local velocity ellipsoids in DM haloes. The
  profiles show the distributions of cosines of the angle ($|\cos\chi|$) between
  the major axis of the local velocity ellipsoid and the halo major
  axis (except of the green dash--dotted profiles for which the angle
  is formed between the major axis of the local velocity ellipsoids
  and the local radial direction). The distributions show that the ellipsoids 
  are preferentially parallel to the halo major axis. Degree of the 
  alignment increases towards the halo centres. The six panels show the
  distribution in $6$ spherical shells of radii as indicated in the
  left upper corners. The black (thick solid), red (solid) and blue
  (dotted) lines show the profiles for different pixel selection: all
  $48$ pixels, $8$ pixels along the major axis and $8$ pixels in the 
  plane perpendicular to the major axis, respectively. The black dashed profiles 
  show the distributions
  inside relaxed haloes. }
\label{sig-orient-pdf}
\end{figure}

Figure~\ref{sig-orient-pdf} shows the distribution of the angle formed
between the major axis of the local velocity ellipsoid and the halo
major axis (black thick profile) or the local radial direction (green
dash--dotted profile). It is clearly seen that the local velocity
ellipsoids in the innermost shells display a strong alignment with the
halo major axis: the black distribution is peaked at $1$, whereas 
the green one is flat. The
alignment is the most prominent in the inner part of the haloes, at
$r<0.2r_{\rm v}$, and gradually vanishes when approaching the virial
sphere. In the three innermost shells (from $0.02r_{\rm v}$ to $0.10r_{\rm v}$), the median angle between the
velocity ellipsoids and the halo major axis are $23$, $23$ and $25$
degrees.

The alignment of the velocity ellipsoids with the halo major axis appears to be stronger along the
major axis than in the perpendicular plane (see the red solid and blue dotted profiles in Figure~\ref{sig-orient-pdf}). 
The difference in degree of alignment between these two regions increases with
radius. In both regions, however, the local velocity ellipsoids
clearly tend to be aligned with the halo major axis. Exceptional
deviation occurs around the virial sphere, where the distribution in
the plane perpendicular to the halo major axis reveals two maxima associated with the alignment
with the local radial directions and the halo major axis (see the
bottom right panel of Figure~\ref{sig-orient-pdf}).

The black dashed profiles in Figure~\ref{sig-orient-pdf} show the
distributions in relaxed haloes. Compared to the sample of all haloes,
one can see that the alignment of the local velocity velocity
ellipsoids with the halo major axis is stronger in the inner parts of
the relaxed haloes. On the other hand, orientations of the velocity
ellipsoids at the virial sphere appear to be distributed in the same
way.

In terms of the alignment of the local velocity ellipsoids, DM 
haloes reveal two distinct zones: the inner part ($r\lesssim
0.2r_{\rm v}$) where the ellipsoids are aligned with the halo major
axis, and the outer part ($r\gtrsim 0.5r_{\rm v}$) where alignment
with the major axis is equally probable as with the radial
direction. The local velocity distribution in the former is consistent
with cylindrical symmetry, whereas in the latter it cannot be
attributed to any simple symmetry such as spherical or
cylindrical. Interestingly, both zones coincide with two
characteristic regimes of the DM density profile: shallower and
steeper than $\rho\propto r^{-2}$.

\subsection{Local anisotropy}

Here, we consider the anisotropy of the local velocity dispersion
tensor. The anisotropy is defined as the ratio of the major-to-minor
velocity dispersions $\sigma_{1}/\sigma_{23}$, where
$\sigma_{23}^{2}=(\sigma_{2}^{2}+\sigma_{3}^{2})/2$.
This definition ignores the fact of triaxiality of the velocity
ellipsoids and, therefore, provides only a partial description of the
ellipsoids. Nevertheless, keeping in mind that the local velocity
ellipsoids are preferentially prolate, i.e. $\sigma_{1}-\sigma_{2}>\sigma_{2}-\sigma_{3}$, we expect
that the $\sigma_{1}/{\sigma}_{23}$ ratio provides the
first order and single-parameter description of the local velocity
distribution (we will consider both the $\sigma_{1}/\sigma_{2}$ and $\sigma_{1}/\sigma_{3}$ ratios in the
following section, where we compare the local and spherically averaged
properties of the velocity dispersion tensor). Contrary to the
definition of the anisotropy parameter $\beta$, here we assume the
cylindrical symmetry of the local velocity distribution.

\begin{figure}
\begin{center}
    \leavevmode
    \epsfxsize=8cm
    \epsfbox[67 60 645 668]{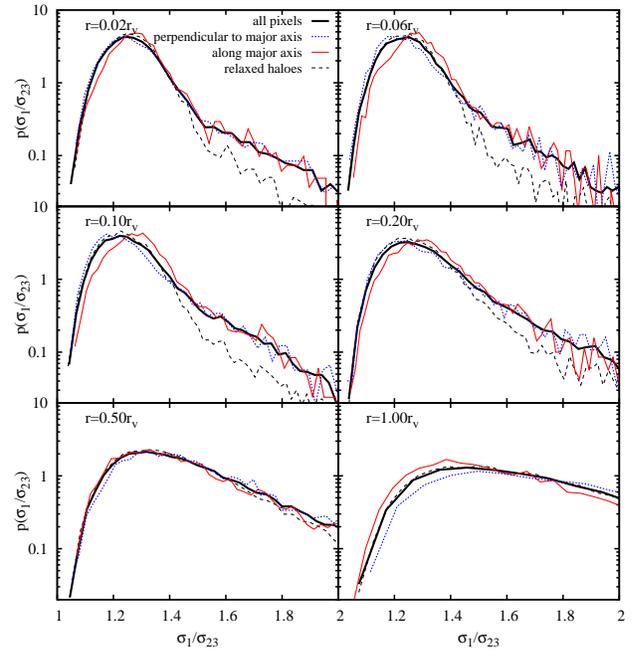}
\end{center}
\caption{Anisotropy $\sigma_{1}/\sigma_{23}$ of the local velocity ellipsoids in DM haloes, where 
$\sigma_{23}^{2}=(\sigma_{2}^{2}+\sigma_{3}^{2})/2$. The six panels show the distributions in $6$ spherical shells of radii 
as indicated in the left upper corners. The colour coding is the same as in Figure~\ref{sig-orient-pdf}. The local 
velocity ellipsoids are highly anisotropic at all radii. Typical value of the anisotropy is $\sigma_{1}/\sigma_{23}\approx1.3$ at 
radii $r<0.5r_{\rm v}$ and increases to $\sigma_{1}/\sigma_{23}\approx1.45$ at the virial radius.}
\label{sig-anisotropy-pdf}
\end{figure}

Figure~\ref{sig-anisotropy-pdf} shows the distribution of the
$\sigma_{1}/\sigma_{23}$ ratio in $6$ spherical shells. The
distributions inside the innermost shells of radii $r\lesssim
0.2r_{\rm v}$ depends barely on radius. The mode is equal to $1.25$ in
the four innermost shells.  The
distributions have long tails extending to high-value
anisotropies. Similar to the case of the velocity dispersion, the
unrelaxed haloes appear to populate preferentially the tails of the
distributions (compare the black thick and dashed profiles
corresponding to the all and relaxed haloes). The anisotropy 
along the major axis of the virial sphere tends to be smaller than in the 
perpendicular plane.

The local velocity ellipsoids at the virial sphere tend to be
substantially more anisotropic than in the inner parts. The mode
occurs at $\sigma_{1}/\sigma_{23}=1.45$ and the distribution
has a slowly decreasing tail at high values. Distributions for relaxed
and unrelaxed haloes appear to be undistinguishable. The anisotropy 
around the halo major axis in these shells tend to be smaller than in the
perpendicular plane, in contrast with the innermost shells (compare the
red solid and blue dotted profiles). In Appendix, we provide an analytic 
model of the velocity anisotropy which recovers the measured profiles shown 
in Figure~\ref{sig-anisotropy-pdf} with $~10$ per cent accuracy.

Similar to the alignment of the velocity ellipsoids, statistical
properties of the $\sigma_{1}/\sigma_{23}$ ratio seem to
differentiate the central parts of the haloes from those around the
virial sphere. The former are characterised by self-similar
distributions with the maximum at approximately $1.25$, whereas the
latter by the distribution gradually getting wider and shifted to
higher values when approaching the virial sphere. The transition
between these two zones is continuous. The approximated radius of the
transition is $0.2r_{\rm v}$ at which the shape of the distribution
starts to deviate from that found in the shells of small radii (see
the middle right panel of Figure~\ref{sig-anisotropy-pdf}).

\section{Spherically averaged anisotropy}

In this section, we address the problem of how to measure the anisotropy
of the velocity dispersion tensor in spherical shells. It is naturally
expected that the velocity dispersion computed in a shell should
represent information contained in the distribution of the local
properties measured in different subvolumes of the shell. 
In particular, spherically averaged profiles (computed in
spherical shells) are expected to recover the most probable values of
the local counterparts.  In order to achieve this, the velocity
dispersion tensor should conform with with the symmetry preferred by
the local velocity ellipsoids. Strong alignment of the velocity
ellipsoids with the halo major axis implies that the velocity
dispersion tensor should be calculated using Cartesian coordinates 
of velocity vectors, as given by the eq.~(\ref{sigma-tensor}). 

\begin{figure}
\begin{center}
    \leavevmode
    \epsfxsize=8cm
    \epsfbox[50 50 590 420]{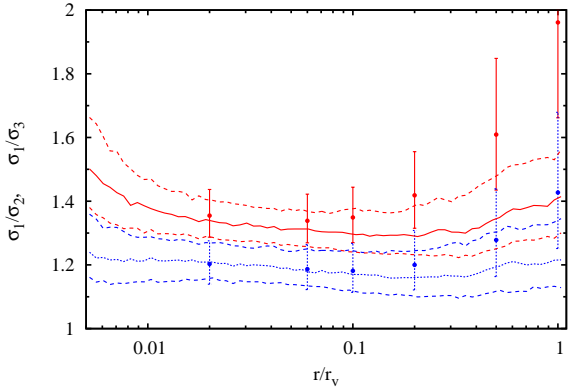}
\end{center}
\caption{Radial profiles of the anisotropy of the velocity dispersion
  tensor calculated in spherical shells. The red solid and blue
  dotted lines are the  profiles for $\sigma_{1}/\sigma_{2}$ and $\sigma_{1}/\sigma_{3}$,
  respectively (with dashed lines showing the first and third quartiles). The
  points with error bars represent the distributions of the local
  values measured in $7$ spherical shells (the median and the error
  bar corresponding to the $50$ per cent probability range). The 
  $\sigma_{1}/\sigma_{2}$ and $\sigma_{1}/\sigma_{3}$ ratios computed 
  in spherical shells reproduce the local values. Discrepancy at radii 
  $r>0.5_{\rm v}$ occurs due to more random orientations of the velocity 
  ellipsoids in this part of the haloes.
  }
\label{ani-shell-min12}
\end{figure}

\begin{figure}
\begin{center}
    \leavevmode
    \epsfxsize=8cm
    \epsfbox[50 50 580 420]{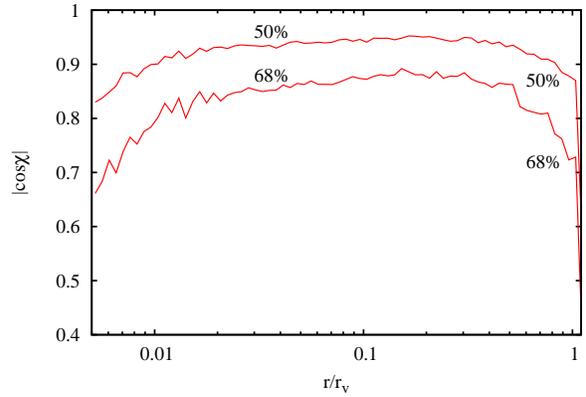}
\end{center}
\caption{Alignment of the velocity ellipsoid calculated in spherical
  shells. The lines show the $50$ (upper) and $68$ (lower) per cent
  probability of the angle ($\cos\chi$) between the major axis of the
  velocity ellipsoid measured in spherical shells and the halo major
  axis. The velocity ellipsoids representing the velocity dispersion 
  tensor calculated in spherical shells are aligned with the halo 
  major axis.}
\label{alignment}
\end{figure}

\begin{figure}
\begin{center}
    \leavevmode
    \epsfxsize=8cm
    \epsfbox[50 50 590 420]{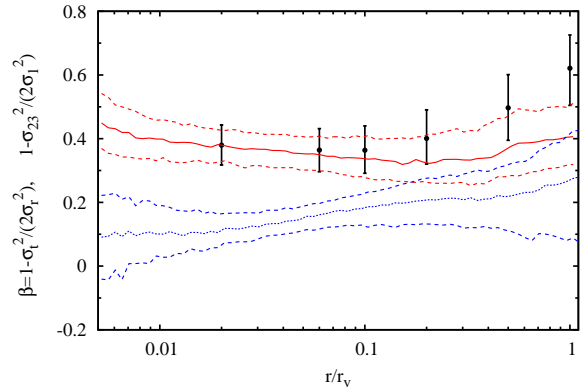}
\end{center}
\caption{Radial profile of the anisotropy of the velocity dispersion
  tensor calculated in spherical shells. The blue dotted and red solid
  profiles show the  anisotropy parameter $\beta=1-(\sigma_{\theta}^{2}+\sigma_{\phi}^{2})/(2\sigma_{r}^{2})$
  and its counterpart defined in cylindrical symmetry,
  i.e. $1-(\sigma_{2}^{2}+\sigma_{3}^{3})/(2\sigma_{1}^{2})$. Black points with error bars
  represent the distributions of the local values measured in $7$
  spherical shells (the median and the error corresponding to the
  $50$ per cent probability range).  The anisotropy defined in cylindrical 
  symmetry and computed in spherical shells recovers the local 
  values. Small discrepancy at radii $r>0.5_{\rm v}$ is due to more random 
  orientations of the velocity ellipsoids in this part of the haloes. The anisotropy 
  parameter $\beta$ is inconsistent with the true anisotropies of the local velocity 
  ellipsoids. Its significantly smaller values result from averaging axially symmetric 
  velocity ellipsoids in spherical 
  coordinate system. Dashed lines show the first and third quartiles. }
\label{anisotropy-shell}
\end{figure}

We used 70 spherical shells with radii equally 
distributed in logarithmic scale between $0.005r_{\rm v}$ and $r_{\rm v}$. 
Figures~\ref{ani-shell-min12}-\ref{anisotropy-shell} show the resulting 
radial profiles of the anisotropy and the orientation of the velocity 
ellipsoids. The profiles show the median and the 50 per cent scatter 
of the values in the halo population. The spherically averaged 
profiles are compared with the local values measured in the shells 
defined in the previous section (points with the error bars showing the median 
and the $50$ per cent scatter).

Figure~\ref{ani-shell-min12} shows the profiles of the 
major-to-median ($\sigma_{1}/\sigma_{2}$) and the
major-to-minor ($\sigma_{1}/\sigma_{3}$) velocity
dispersions. The spherically averaged profiles of the anisotropy are fairly
consistent with the local values at radii $r<0.5r_{\rm v}$. In this range of
radii, the anisotropy is well-approximated by a flat profile with
$\sigma_{1}/\sigma_{2}\approx1.25$ and $\sigma_{1}/\sigma_{3}\approx1.42$.

The spherically averaged anisotropy appears to underestimate the true
local values at radii $r>0.5r_{\rm v}$. This is particularly well
visible for the $\sigma_{1}/\sigma_{3}$ ratio, for which deviation between 
the median of the local and spherically averaged values of the anisotropy 
is of the order of $2\sigma$ at $r\approx r_{\rm v}$. This
discrepancy is mostly caused by a tension between the assumed symmetry
of the velocity ellipsoids and the true orientations of the local
velocity ellipsoids. Larger scatter in the alignment of the local
velocity dispersions (more random orientations) produces artificially
more isotropic velocity dispersion tensor calculated in spherical
shells.

Figure~\ref{alignment} shows the profiles of the angle between the major
axis of the velocity ellipsoid measured in spherical shells and the
halo major axis. The lines indicate the $50$ and $68$ per cent
probability range of the angle measured at every radius (with the most
probable values at all radii equal to $|\cos\chi|=1$). This clearly
shows that the velocity ellipsoids computed in spherical shells are
aligned with the halo major axis. The scatter of the angles is
comparable to the scatter of the local values (compare with
Figure~\ref{sig-orient-pdf}).

For the sake of illustration how $\beta$ parameter misrepresents the
true picture of velocity anisotropy in DM haloes, we compare in
Figure~\ref{anisotropy-shell} the spherically averaged profiles of
$\beta$ and its analogue measured in a Cartesian coordinates system aligned 
with the semi-principle axes of the halo shape (cylindrical symmetry), 
i.e.  $1-(\sigma_{2}^{2}+\sigma_{3}^{2})/(2\sigma_{1}^{2})$.  As expected,
the $\beta$ profile differs substantially from its counterpart
calculated in cylindrical symmetry.  In particular, nearly
isotropic velocity distribution (in spherical coordinate system) in
the halo centre, i.e. $\beta\approx0$, is an effect of angular
averaging over velocity ellipsoids which are strongly aligned with the
halo major axis. The true orientations of the local velocity
ellipsoids violate some crucial assumptions underlying the definition
of the anisotropy parameter $\beta$ (radial orientations of the local
velocity ellipsoids). Violation of these assumptions leads to
artificially more isotropic velocity dispersion tensor (in terms of
$\beta$).

In Figure~\ref{anisotropy-shell}, we also compare the spherically
averaged profiles with the local values. The anisotropy defined in
cylindrical symmetry recovers the local values at $r<0.5r_{\rm v}$. As
in the case of Figure~\ref{ani-shell-min12}, the discrepancy at large
radii is due to more random orientations of the local velocity
ellipsoids. On the other hand, the anisotropy parameter $\beta$ gives
erroneous impression of significantly more isotropic velocity
distribution. The median $\beta$ profile lies $(2-4)\sigma$ lower than
the median profile of the true local anisotropy. Using $\beta$
parameter leads to artificial isotropisation of the velocity
distribution not only in the centres of DM haloes, but also at radii
comparable to the virial radius.

The anisotropy parameter $\beta$ leads to a misleading picture of the
true anisotropy of the velocity dispersion tensor not only when
applying to spherical shells, but also locally. For example, the
median $\beta$ in the equatorial pixels of the innermost shells (see
Figure~\ref{beta-density}) point to nearly isotropic or weakly
tangentially-biased velocity dispersion tensor, i.e. $\beta \lesssim
0$. In fact, this is a consequence of a peculiar orientation of the
local velocity ellipsoids whose major axes are preferentially parallel
to the direction of the polar angle in the equatorial
stripe. Averaging over the angles, a commonly used routine based on
adopting the mean $(\sigma_{\theta}^{2}+\sigma_{\phi}^{2})/2$ as the
variance of the tangential velocity, mixes the major and minor axes of
the velocity ellipsoids and thus gives rise to an artificial
isotropisation.

\section{Discussion and conclusions}

We studied the properties of the local velocity dispersion tensors in
cluster-size simulated DM haloes. We found that the velocity
ellipsoids representing the tensors are strongly aligned with the halo
major axis defined as the axis minimising the moment of inertia
calculated inside the virial sphere. Statistical properties of the
orientations and anisotropies of the local velocity dispersion tensor
do not vary with radius in the central parts of the haloes
($r<0.2r_{\rm v}$). At large radii, the orientations become gradually
randomised and the ellipsoids more elongated. These two distinct zone
of the haloes coincide with two characteristic regimes of the DM
density profile: shallower and steeper than $\rho\propto r^{-2}$.

Alignment of the local velocity ellipsoids is inconsistent with
spherically symmetric velocity distribution (radially oriented
ellipsoids) that is an assumption underlying definition of the
anisotropy parameter $\beta$. As a consequence, using the anisotropy
parameter $\beta$ (assuming radial orientations of the local velocity
ellipsoids) leads to an erroneous picture of significantly more
isotropic velocity distributions than they really are. Typical ratio
of the major-to-minor axis of the local velocity ellipsoids is equal
to $1.3$ at radii $r<0.5r_{\rm v}$ and increases strongly with radius
in the outer part of the haloes. In the light of the results presented in this
paper, the assumption of spherically symmetric velocity distributions
in DM haloes acts as an artificial phase mixing which may lead to a
picture, in which DM haloes appear to be more equilibrated than they
are. This phase mixing is one of the factors underlying 
the linear relation between $\beta$ and the logarithmic density slope 
found in simulated DM haloes \citep{Han06}.

Similar trends of the $\beta$ profiles with the angle from the halo major 
axis were found in controlled simulations of halo mergers \citep{Spa12}. 
This resemblance suggests that cylindrical symmetry of the velocity 
distributions in cosmological haloes is a remnant of major mergers.

In order to capture physical properties of the velocity anisotropy in 
DM haloes, we recommend to use a parameter based on the ratio 
of the velocity dispersions along and perpendicular to the halo major 
axis. A natural analogue of the classical $\beta$ parameter is
\begin{equation}
\beta_{\rm cyl}=1-\frac{\sigma_{2}^{2}+\sigma_{3}^{2}}{2\sigma_{1}^{2}},
\end{equation}
where $\sigma_{1}$, $\sigma_{2}$ and $\sigma_{3}$ are the velocity 
dispersions along the major, medium, and minor axis of the halo shape 
ellipsoid. This parameter describes the velocity anisotropy in cylindrical 
symmetry with respect to the halo major axis that is consistent 
with the orientations of the local velocity ellipsoids. An analytic model 
of the velocity anisotropy is provided in Appendix.

\subsection{Dynamical modelling of quasi-spherical systems}

Although our studies are based on the specific sample of DM haloes
(cluster--size haloes), the overall picture of the velocity ellipsoids
aligned with the halo major axis may likely be generic. 
The argument supporting this expectation is twofold. First, the
alignment is more prominent in relaxed haloes rather than in recent
mergers. Second, all processes of anisotropic halo formation are
common to all DM haloes and, therefore, the same mechanisms breaking
spherical symmetry of the velocity distribution should operate at all
scales.

Our results may have some consequences for the mass modelling of
kinematical data in such objects as galaxy clusters, elliptical
galaxies and dwarf spheroidals. Most of dynamical models
assume spherically symmetric velocity distributions whose anisotropy
is quantified with the $\beta$ parameter. It has barely ever been
verified whether this assumption conforms with the true orbital
structure inside these objects.  This problem was addressed by
\citet{Cap07} in a detailed study of the stellar kinematics in massive
ellipticals. It was shown that the velocity ellipsoids are
preferentially aligned with the galaxy shape, on the contrary to what
is assumed when considering $\beta$ parameter. Some signatures of
similar orientation of the velocity ellipsoids in galaxy clusters were
also demonstrated by \citet{Ski12}.

Alignment of the local velocity ellipsoids with the major axis may
inevitably affect the mass inference with the use of dynamical models
assuming radial orientations of the ellipsoids. An additional analysis
is required to address this question in a fully quantitative way.  However, in
order to introduce the scale of the problem, we note that, in the
light our results, the projected velocity dispersions in two
perpendicular directions may differ by as much as $50$ per cent
(see Figure~\ref{ani-shell-min12}). Using the $M_{\rm v}\propto
\sigma^{3}$ scaling relation, this leads to the ratio of the mass
estimates in both projections equal to $3.4$.

\subsection{Observational constraints on $\beta$}

The anisotropy parameter $\beta$ has been measured for a number of
astrophysical systems. The estimates obtained from the stacked
kinematical data of galaxies in clusters \citep{Biv04,Woj10} and the
satellite galaxies around isolated hosts \citep{Woj12} are fairly
compatible with the spherically averaged $\beta$ profiles from the
simulations. We emphasise, however, that this consistency does not
imply the fact that $\beta$ describes the true orbital anisotropy,
because stacking kinematical data introduces artificial
sphericalisation of the phase-space distribution that is equivalent to
measuring $\beta$ in spherical shells of simulated objects.

Observational constraints on the spherical anisotropy $\beta$ in
individual clusters exhibit a huge variety of profiles
\citep{Ben06,Hwa08,Woj10,Woj09}.  Some measurements are far outside
the margins allowed by the spherically averaged profiles of simulated
clusters \citep[see e.g.][]{Hwa08}.  We suspect that this substantial
scatter of observational constraints on $\beta$ in individual clusters
may result from ignoring the angle between the line of sight and the
major axis of the velocity ellipsoids. This projection effect needs to
be addressed in future in a quantitive analysis of mock kinematical
data generated from cosmological simulations.

\section*{Acknowledgments}
The Dark Cosmology Centre is funded by the Danish National Research Foundation. 
We gratefully acknowledge the anonymous referee for constructive comments. 
RW thanks Gary Mamon, Jens Hjorth, Steen Hansen, Martin Sparre and Andreas Skielboe 
for critical reading of the manuscript and insightful comments.  RW is also 
grateful to Noam Libeskind for fruitful discussions during the CLUES meeting 
in Lyon. AK acknowledges support of NSF grant AST-1009908 to NMSU. 
SG acknowledges the funding of the collaboration with AK by DAAD. 
Database used in this paper and the web application providing online
access to it were constructed as part of the activities of the German
Astrophysical Virtual Observatory as result of a collaboration between
the Leibniz-Institute for Astrophysics Potsdam (AIP) and the Spanish
MultiDark Consolider Project CSD2009-00064.  The Bolshoi simulation
was run on the NASA's Pleiades supercomputer at the NASA Ames Research
Center.

\bibliography{master}

\appendix
\section{Analytic model of the velocity anisotropy}
The following simplified model captures the main features of the
velocity anisotropy studied in this work. In the reference frame whose axes are aligned along the 
principle axes of the density distribution (with $x$, $y$ and $z$ axes corresponding 
to the major, medium and minor axis, respectively), the 
number-density $n$ as the function of velocities $\vec V=\{v_x,v_y,v_z\}$ 
and coordinates $\vec r=\{x,y,z\}$ can be written as :
\begin{eqnarray}
n(\vec V,\vec r) &=& \frac{n_0(\vec r)}{(8\pi^3\sigma^2_x\sigma^2_y\sigma^2_z)^{1/2}}\times \nonumber\\
         && \exp\left[-\frac{v_x^2}{2\sigma^2_x}-\frac{v_y^2}{2\sigma^2_y}-\frac{v_z^2}{2\sigma^2_z}\right],\\
     \beta_y(x,y,z) &\equiv& 1-\frac{\sigma_y^2}{\sigma^2_x} = \beta_{0y}\left(1-\frac{w_y^2(0.4+w)}{1+w}\right), \\
     \beta_z(x,y,z) &\equiv& 1-\frac{\sigma_z^2}{\sigma^2_x} =\beta_{0z}\left(1-\frac{w_z^2(0.4+w)}{{1+w}}\right), \\
     w_y &=& \left(\frac{y}{r_{sy}w}\right), \quad
     w_z = \left(\frac{z}{r_{sz}w}\right),  \\
     w^2 &=& \left(\frac{x}{r_{sx}}\right)^2 +\left(\frac{y}{r_{sy}}\right)^2 +\left(\frac{z}{r_{sz}}\right)^2,
\end{eqnarray}
where $n_0(x,y,z)$ is the density profile of the halo and $r_{sx}$,$r_{sy}$ and $r_{sz}$ are 
the scale radii of the NFW density profile \citep{Nav97} fitted along the major, medium and minor axis, respectively. 
For cluster--size haloes studied in the paper 
\begin{equation}
 \beta_{0y} \approx 0.35, \quad  \beta_{0z} \approx 0.45
\end{equation}
The model assumes a Gaussian distribution of velocities with the local velocity dispersion tensor aligned with the 
semi-principle axes of the halo. The ratios of the medium-to-major and minor-to-major velocity dispersions are 
functions of the position, $\beta_y(x,y,z)$ and $\beta_z(x,y,z)$ respectively. The velocity dispersion along a given 
semi-principle axis of the halo is a free 
function of the model which may be determined using the Jeans equation. The model recovers the 
measurements of the velocity anisotropy shown in Figures~\ref{sig-anisotropy-pdf},\ref{ani-shell-min12} 
and \ref{anisotropy-shell} with $~10$ per cent accuracy.

\end{document}